\documentclass[10pt,conference,final,twocolumn]{IEEEtran}
\usepackage[top=0.71in,bottom=1.04in,left=0.625in,right=0.625in]{geometry}

\usepackage{amsfonts}
\usepackage[cmex10]{amsmath}
\allowdisplaybreaks[4]  
\usepackage{amsthm}
\usepackage{amssymb}
\usepackage{mdwmath}
\usepackage{mathtools}
\usepackage{cases}
\usepackage{empheq} 
\usepackage{bm}

\theoremstyle{definition}

\usepackage{tabularx}
\usepackage{array}
\usepackage{mdwtab}
\usepackage{multirow}
\usepackage{booktabs}  
\usepackage{diagbox} 
\usepackage{makecell}
\newcolumntype{C}[1]{>{\centering\arraybackslash}p{#1}}

\usepackage{textcomp}
\def\BibTeX{{\rm B\kern-.05em{\sc i\kern-.025em b}\kern-.08em
T\kern-.1667em\lower.7ex\hbox{E}\kern-.125emX}}

\usepackage{graphicx}
\usepackage{xcolor}
\usepackage{caption}  
\usepackage[labelformat=simple,font=footnotesize]{subcaption}
\graphicspath{ {./img/} }

\usepackage{cite}

\usepackage{makeidx}

\usepackage{algorithm}
\usepackage{algorithmic}

\begin{document}
\bstctlcite{IEEEexample:BSTcontrol}
\title{
AWM: An All-in-One Wireless Foundation Model for Multi-User Channel Acquisition and Localization Without Downstream Tuning
}
\author{Zirui~Chen\textsuperscript{$\dagger$}, 
Ziqing~Xing\textsuperscript{$\dagger$}\textsuperscript{$\ddagger$}, 
Zhaoyang~Zhang\textsuperscript{$\dagger$}\textsuperscript{$\ddagger$}, Hongning~Ruan\textsuperscript{$\dagger$}\textsuperscript{$\ddagger$}, 
Zili~Wang\textsuperscript{$\dagger$}, 
and Zhaohui~Yang\textsuperscript{$\dagger$}\\ 
\IEEEauthorblockA{
    \textsuperscript{$\dagger$}College of Information Science and Electronic Engineering, Zhejiang University, Hangzhou 310027, China\\
    \textsuperscript{$\dagger$}Zhejiang Provincial Laboratory of Multi-Modal Communication Networks \\and Intelligent Information Processing, Hangzhou 310027, China\\
    \textsuperscript{$\ddagger$}Institute of Fundamental and Transdisciplinary Research, Zhejiang University, Hangzhou 310058, China\\
    E-mails: \{ziruichen, ziqing\_xing, ning\_ming, rhohenning, zili\_wang, yang\_zhaohui\}@zju.edu.cn \vspace{-1.5em}
}
}
\maketitle
\begin{abstract}
Wireless foundation models (WFMs) seek to unify diverse wireless tasks and scenarios, yet existing methods still rely on task- or scenario-specific components or deployment-time fine-tuning.
This paper presents an all-in-one wireless model (AWM), a WFM that unifies multi-user channel acquisition and localization, despite their distinct data modalities and functional objectives.
We reveal that user location and channel state information (CSI) at different frequencies can be viewed as attribute-specific measurements of the same wireless link. 
By introducing in-scenario same-attribute reference pairs as prompts, AWM converts heterogeneous attributes into alignable relativity representations, formulating intra-band channel interpolation, inter-band channel prediction, and wireless localization as unified wireless-state transformations. 
Further, a multi-task MateFormer is developed to perform source-attribute relativity extraction, cross-attribute relativity alignment, and target-attribute synthesis within a unified architecture, avoiding task-specific output heads or downstream fine-tuning. 
Comprehensive experiments demonstrate the effectiveness of model unification, including substantial gains from cross-task knowledge sharing, seamless generalization to unseen scenarios, and an efficient closed-loop service for mobile users.
\end{abstract}

\section{Introduction}\label{introduction}

As the sixth-generation (6G) communication systems evolve, artificial intelligence (AI) is expected to empower essential capabilities such as channel state information (CSI) acquisition and user state sensing.
However, conventional wireless AI schemes are mostly tailored for specific scenarios and tasks, making them difficult to generalize across massive heterogeneous use cases \cite{chen2024wBAIM}. 
Whether wireless models can be unified has remained a persistent question, also motivating recent research on wireless foundation models (WFMs).

Masked channel modeling is a widely studied pre-training method, 
which attempts to learn unified channel features from large-scale CSI datasets while improving channel-representation tasks such as estimation and prediction \cite{guo2026maskHmodel}.
With task-specific parameters and modules, these learned representations can then be adapted to downstream applications such as localization and beamforming \cite{liu2026wifo2}.
Other pre-training paradigms include cross-domain and cross-modal contrastive learning \cite{jiang2025CSICLIP,liu2026wifoMiSAC}.
However, these methods typically require task-specific output heads to handle different tasks, 
leaving a gap between existing WFMs and the all-in-one vision. This still entails laborious model development and adaptation and obscures whether the resulting functionality is primarily determined by the shared backbone or the adaptation modules.

In parallel with downstream fine-tuning, some studies have introduced in-scenario reference context, directly or indirectly opening a path toward unified models across multiple functionalities.
First, channel deduction \cite{chenCD2025} combines current coarse estimates with historical CSI, 
substantially improving the accuracy and reliability of channel acquisition while unifying channel prediction and estimation within a common framework.
Its extensions, including spatial channel deduction (SCD) \cite{chen2025SCD} and geometry-aided channel deduction (GCD) \cite{ruan2026GCD}, extend the reference scope from temporal to spatial neighborhoods and improve channel representation by exploiting available environmental information.
Most notably, for environment-dependent wireless localization, researchers retrieve in-scenario reference samples based on CSI fingerprint similarity or coarse spatial neighborhoods, and construct cross-scenario localization models via analogical learning (AL) \cite{chen2025ALLoc} or in-context learning (ICL) \cite{pan2026RA-LWLM}.
Although these studies mainly focus on performance gains and scenario-level generalization, without achieving complete unification across multiple tasks and frequencies, their advanced architectures have brought wireless models increasingly close to an all-in-one design.

Building on these studies, we propose an all-in-one wireless model (AWM).
AWM inherits the cross-scenario generalization of AL \cite{chen2025ALLoc} and jointly supports intra-band interpolation, inter-band prediction, and wireless localization, which span bands and modalities.
It also accommodates arbitrary channel frequencies, allowing multiple users to access flexibly.
The key insight is that, although user locations and CSI at different frequencies exhibit distinct feature structures and numerical characteristics, they are all attribute-specific state measurements of the same link.
Analogy with same-attribute references yields a dimensionless relative representation that suppresses attribute-specific differences and aligns link features.
Accordingly, we redefine these tasks as transformations from a source attribute to a target attribute conditioned on references of the same paired attributes.
The source and target attributes correspond to the model input and output, respectively.
Retrieved from the user's spatial neighborhood, these reference pairs convey task information in a few-shot manner.
Thus, different tasks can be unified to a single model architecture and the parameters during training and deployment, providing the following benefits:
\begin{itemize}
    \item Sharing knowledge across tasks and frequencies, yielding substantial performance gains, with inter-band prediction in particular developing an emergent capability from being unusable to meeting practical requirements.
    \item Increasing the information density of limited training data, leading to clear and unsaturated scaling behavior, leaving room for further gains in wireless AI performance.
    \item Reducing the number of required models that grows linearly with the tasks, scenarios, and frequencies to only one, while enabling a closed-loop service through coordinating internal functions.
\end{itemize}

\section{System Requirements and Technical Basis} \label{sec:system_and_basis}

\subsection{System Framework and Functional Requirements} \label{subsec:system}

\begin{figure}[!htbp]
\centering
\vspace{-0.5em}
\includegraphics[width=\linewidth]{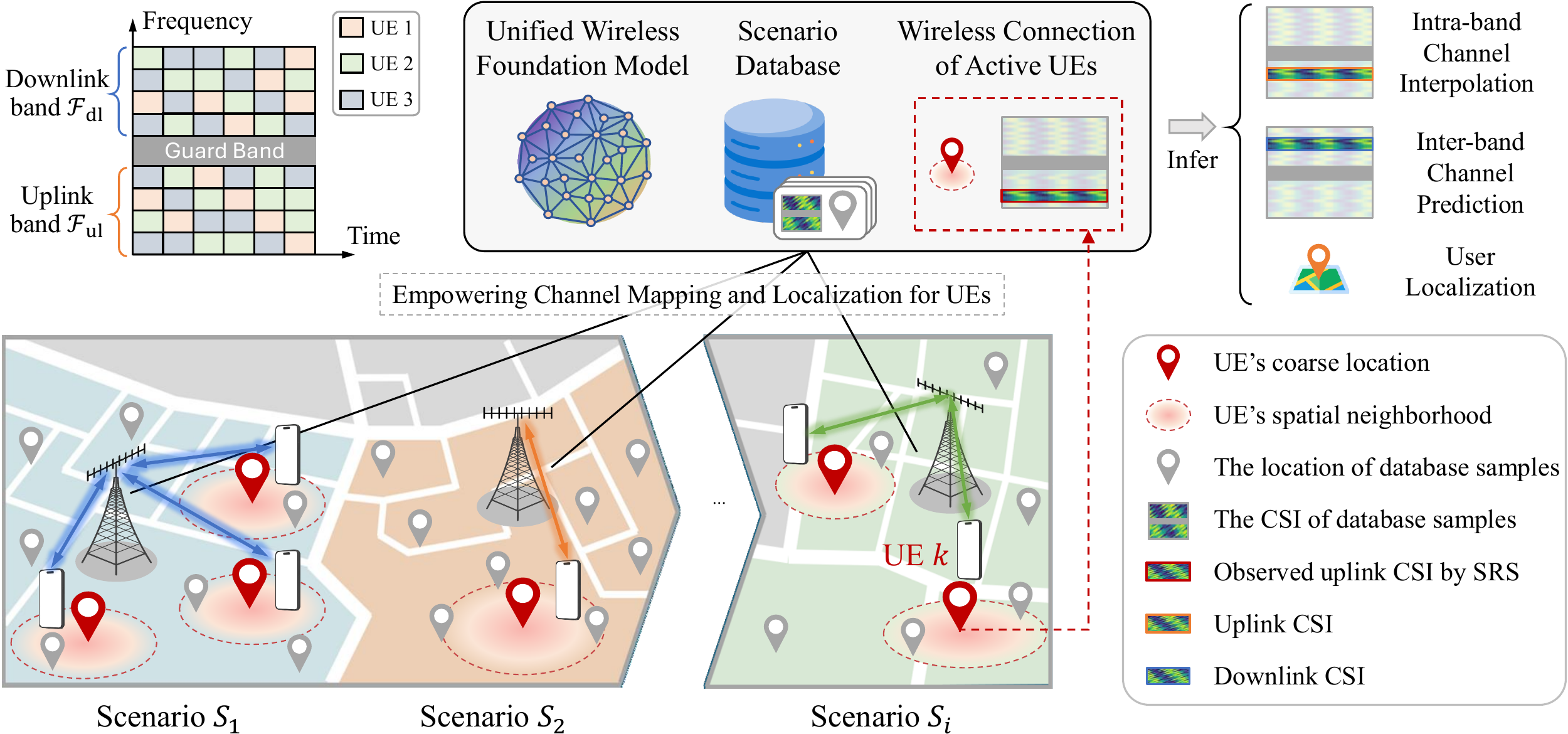}
\caption{The consideted OFDMA communication system across multiple scenarios.}
\label{fig:system}
\vspace{-1em}
\end{figure}

We consider a cellular system comprising multiple wireless scenarios, where the $i$-th scenario is denoted by $S_i$, as illustrated in Fig. \ref{fig:system}. 
In each scenario, a frequency-division duplexing multiple-input multiple-output (FDD-MIMO) system is considered, with the uplink and downlink frequency bands denoted by $\mathcal F_{\mathrm{ul}}$ and $\mathcal F_{\mathrm{dl}}$, respectively.
The base station (BS), equipped with $N_{\rm t}$ antennas, serves multiple single-antenna user equipments (UEs) via orthogonal frequency-division multiple access (OFDMA).
In scenario $S_i$, the location of UE $k$ is denoted by $\mathbf{x}_k^i\in\mathbb{R}^2$, and its channel at frequency $f$ is represented by $\mathbf{h}_k^i(f)\in\mathbb{C}^{N_{\rm t}}$.
Under constraints on uplink pilot overhead and transmit power, UE $k$ transmits a sounding reference signal (SRS) only on its assigned uplink sounding carrier $f_{\mathrm{ul}}^{\mathrm{SRS}}[k]\in\mathcal F_{\mathrm{ul}}$, from which the BS estimates the uplink CSI $\mathbf{h}_k^i(f_{\mathrm{ul}}^{\mathrm{SRS}}[k])$.
In addition, a meter-level coarse location $\widetilde{\mathbf{x}}_k^i$ is 
available from a global navigation satellite system (GNSS) or previous-slot localization. 
Based on this coarse location, it can retrieve reference samples spatially adjacent to UE $k$ from the database of scenario $S_i$ to assist its wireless tasks, and denote their index set by $\Theta_k^i$ \cite{chen2025ALLoc}.

To support functionalities such as uplink transmission, downlink precoding, and user localization, the BS is expected to perform the following tasks based on the measured SRS channel:
(i) intra-band channel interpolation $\mathbf h_k^i(f_{\mathrm{ul}}^{\mathrm{SRS}}[k]) \mapsto \mathbf h_k^i(f_{\mathrm{ul}}[k])$;
(ii) inter-band channel prediction $\mathbf h_k^i(f_{\mathrm{ul}}^{\mathrm{SRS}}[k]) \mapsto \mathbf h_k^i(f_{\mathrm{dl}}[k])$; and
(iii) wireless localization $\mathbf h_k^i(f_{\mathrm{ul}}^{\mathrm{SRS}}[k]) \mapsto \mathbf x_k^i$.
Here, $f_{\mathrm{ul}}[k]\in\mathcal F_{\mathrm{ul}}\setminus\{f_{\mathrm{ul}}^{\mathrm{SRS}}[k]\}$ and $f_{\mathrm{dl}}[k]\in\mathcal F_{\mathrm{dl}}$ denote the uplink and downlink carriers assigned to UE $k$, respectively.
For unified modeling, the above three tasks can be categorized into two fundamental functionalities: channel mapping and wireless localization.
We define the source frequency as $f_{\rm s}=f_{\mathrm{ul}}^{\mathrm{SRS}}[k]$, and the target frequency $f_{\rm t}$ corresponds to $f_{\mathrm{ul}}[k]$ or $f_{\mathrm{dl}}[k]$ depending on the task requirement.
Accordingly, these two functionalities are formulated as
\vspace{-0.6em}
\begin{equation}\label{eq:functional_requirements}
\vspace{-0.4em}
\mathbf h_k^i(f_{\rm s})\mapsto
\begin{cases}
\mathbf h_k^i(f_{\rm t}), & \text{channel mapping},\\
\mathbf x_k^i, & \text{wireless localization}.
\end{cases}
\end{equation}
Although these tasks span different carrier frequencies and data modalities, channel mapping and wireless localization share some common physical basis.
Within a given scenario, the UE location determines the geometry of the wireless link, while the CSI at different frequencies characterizes its electromagnetic (EM) responses. 
Therefore, the location and the frequency-dependent CSI can be regarded as different physical attributes of the same underlying wireless link. 
This commonality provides the foundation for jointly supporting channel acquisition and wireless localization within a unified learning framework.

\subsection{Spatial Channel Deduction (SCD)} \label{subsec:SCD}

As an emerging channel acquisition approach, SCD \cite{chen2025SCD} jointly utilizes the coarse channel estimate of an UE and the reference channels in its spatial neighborhood to improve performance under limited pilot overhead.
Since nearby locations generally share similar propagation conditions, these reference channels complement the large-scale multipath features that cannot be sufficiently captured by the coarse estimate.

The original SCD is designed for two-dimensional space-frequency channel acquisition.
For UE $k$ in scenario $S_i$, let $\mathbf{H}_k^i\in\mathbb{C}^{N_{\rm t}\times N_{\rm c}}$ denote its complete channel and $\mathbf{H}_k^i[\Omega]$ the coarse estimate observed at the pilot indices $\Omega$.
SCD learns the following deduction function:
\vspace{-0.3em}
\begin{equation}\label{eq:original_SCD}
\vspace{-0.3em}
\mathrm{NN}_{\mathrm{SCD}}
\left(
\{\mathbf{H}_j^i\}_{j\in\Theta_k^i},
\mathbf{H}_k^i[\Omega]
\right)
\mapsto
\widehat{\mathbf{H}}_k^i.
\end{equation}
In this formulation, $\mathbf{H}[\Omega]$ is a subset of $\mathbf{H}$. 
Accordingly, \eqref{eq:original_SCD} can be rewritten as
\vspace{-0.3em}
\begin{equation}\label{eq:paired_SCD}
\vspace{-0.3em}
\mathrm{NN}_{\mathrm{SCD}}
\left(
\{(\mathbf{H}_j^i[\Omega],\mathbf{H}_j^i)\}_{j\in\Theta_k^i},
\mathbf{H}_k^i[\Omega]
\right)
\mapsto
\widehat{\mathbf{H}}_k^i.
\end{equation}
This reformulation preserves the information content of both the input and the output.
It therefore defines a learning task equivalent to the original SCD for the neural network.
In this formulation, $\mathbf{H}[\Omega]$ and $\mathbf{H}$ can be regarded as the source and target channel domains, respectively.
The neighboring channels $\mathbf{H}_j^i$ thus form source--target reference pairs, i.e., $(\mathbf{H}_j^i[\Omega], \mathbf{H}_j^i)$.

Under the system model considered in this paper, channel acquisition aims to infer the CSI at target frequency $f_{\rm t}$ from the uplink CSI at source frequency $f_{\rm s}$. Accordingly, SCD is extended as
\vspace{-0.3em}
\begin{equation}\label{eq:extended_SCD}
\vspace{-0.3em}
\mathrm{NN}_{\mathrm{SCD}}
\left(
\{(\mathbf{h}_j^i(f_{\rm s}),\mathbf{h}_j^i(f_{\rm t}))\}_{j\in\Theta_k^i},
\mathbf{h}_k^i(f_{\rm s})
\right)
\mapsto
\widehat{\mathbf{h}}_k^i(f_{\rm t}).
\end{equation}
Here, $\mathbf{h}(f_{\rm s})$ is no longer a subset of $\mathbf{h}(f_{\rm t})$. 
The two channels may correspond to different carriers within the uplink band or belong to the uplink and downlink bands, respectively.
This extension only modifies the source and target channel domains involved in SCD, while preserving its core principle: spatial reference channels capture the local propagation environment of the UE and thereby facilitate the target channel recovery.

\subsection{Analogical Learning (AL)-Based Wireless Localization} \label{subsec:ALLoc}

Wireless localization aims to estimate the UE location from its CSI. 
However, inherent differences in propagation environments across multiple scenarios make it difficult to learn a unified mapping $\mathbf{h}\mapsto\mathbf{x}$.
AL \cite{chen2025ALLoc} addresses this problem by introducing channel--location reference pairs from the spatial neighborhood of the UE.

Under the system model considered in this paper, AL-based wireless localization is formulated as
\vspace{-0.35em}
\begin{equation}\label{eq:AL_localization}
\vspace{-0.35em}
\mathrm{NN}_{\mathrm{AL}}
\left(
\{(\mathbf{h}_j^i(f_{\rm s}),\mathbf{x}_j^i)\}_{j\in\Theta_k^i},
\mathbf{h}_k^i(f_{\rm s})
\right)
\mapsto
\widehat{\mathbf{x}}_k^i.
\end{equation}
Owing to the similar propagation conditions within the neighborhood, the reference-pair set $\{(\mathbf{h}_j^i(f_{\rm s}),\mathbf{x}_j^i)\}_{j\in\Theta_k^i}$ implicitly embeds the scenario information of UE $k$.
Furthermore, \cite{chen2025ALLoc} proposed the MateFormer model to efficiently implement the AL process in \eqref{eq:AL_localization}.
It first extracts the channel-domain relativity between the UE's channel $\mathbf{h}_k^i(f_{\rm s})$ and the reference channels $\{\mathbf{h}_j^i(f_{\rm s})\}_{j\in\Theta_k^i}$, 
aligns this relativity to spatial domain, 
and then synthesizes $\widehat{\mathbf{x}}_k^i$ from the reference locations $\{\mathbf{x}_j^i\}_{j\in\Theta_k^i}$. 
In this way, AL converts the scenario-dependent channel--location mapping into reference-pair-based relativity inference, enabling a unified localization model across diverse scenarios.

\section{Proposed All-in-One Wireless Model (AWM)} \label{sec:AWM}

\subsection{Relativity Representation is All You Need} \label{subsec:relativity}

Based on the physical commonality discussed in Section~\ref{subsec:system}, two functionalities in \eqref{eq:functional_requirements} can be viewed as homogeneous wireless state transformations, both inferring a target state of the same wireless link from observed CSI.

At the technical level, SCD and AL also exploit scenario information in a common form: both retrieve reference samples from the spatial neighborhood of the UE to assist inference.
These reference samples experience propagation conditions similar to those of current UE and thus implicitly embed the local EM environment information, serving as scenario-specific context for channel mapping or wireless localization.
According to \eqref{eq:extended_SCD} and \eqref{eq:AL_localization}, SCD and AL can be unified as
\vspace{-0.5em}
\begin{equation}\label{eq:unified_state_transformation}
\vspace{-0.2em}
\mathrm{NN}
\left(
\{(\mathbf A_j^i,\mathbf B_j^i)\}_{j\in\Theta_k^i},
\mathbf A_k^i
\right)
\mapsto
\widehat{\mathbf B}_k^i.
\end{equation}
Here, $\mathbf A$ and $\mathbf B$ denote abstract representations of the attributes in the source and target domains, respectively; $\{(\mathbf A_j^i,\mathbf B_j^i)\}_{j\in\Theta_k^i}$ denotes the neighborhood reference source-target pairs, and $\mathbf A_k^i$ is the source attribute provided by the current UE.
In this paper, the source attribute $\mathbf A$ is always the uplink CSI $\mathbf h(f_{\rm s})$ acquired through the SRS.
When the target attribute $\mathbf B$ is the CSI $\mathbf h(f_{\rm t})$ at the target frequency, \eqref{eq:unified_state_transformation} corresponds to channel mapping, whereas it corresponds to wireless localization when $\mathbf B$ is the UE location $\mathbf x$.
Therefore, \eqref{eq:unified_state_transformation} provides a unified formulation for wireless-state transformation between divese attributes.

Nevertheless, \eqref{eq:unified_state_transformation} lacks a guiding mechanism for cross-attribute transformation and alignment, which can hinder learning with limited training samples \cite{chen2025ALLoc}.
Therefore, we reemploy AL, which was originally designed for localization, to general wireless state transformation tasks and implement \eqref{eq:unified_state_transformation} as
\vspace{-0.2em}
\begin{equation}\label{eq:relativity_state_transformation}
\vspace{-0.2em}
\mathrm{NN}_{\mathrm{use\_rela}}
\left(
\mathrm{NN}_{\mathrm{ext\_rela}}\left(\mathbf A_j^i,\mathbf A_k^i\right),
\mathbf B_j^i
\right)_{j\in\Theta_k^i}
\mapsto
\widehat{\mathbf B}_k^i.
\end{equation}
In \eqref{eq:relativity_state_transformation}, $\mathrm{NN}_{\mathrm{ext\_rela}}$ first extracts the relativity representation between the source attribute $\mathbf A_k^i$ of the current UE and the reference source attributes $\{\mathbf A_j^i\}_{j\in\Theta_k^i}$.
Then, based on the correspondence between $\mathbf A_j^i$ and $\mathbf B_j^i$ in each reference pair, $\mathrm{NN}_{\mathrm{use\_rela}}$ aligns this relativity with the target attribute domain and synthesizes $\widehat{\mathbf B}_k^i$ from the target attributes $\{\mathbf B_j^i\}_{j\in\Theta_k^i}$ of reference samples.
This design uses relativity representations to learn how the local scattering environment affects current samples through EM interactions and to infer target attributes, while the dimensionless nature of relativity also facilitates alignment and transformation across attributes during network optimization.
This unifies diverse wireless modalities and tasks under a common and feasible learning objective.
It avoids separate model development and allows data, knowledge, and training costs to be shared across tasks and scenarios.
The principle of ``relativity representation is all you need'' thus provides the basis for the all-in-one design of AWM.

\subsection{The Learning Architecture of AWM} \label{subsec:architecture}

\begin{figure*}[t]
  \centering
  \includegraphics[width=0.95\linewidth]{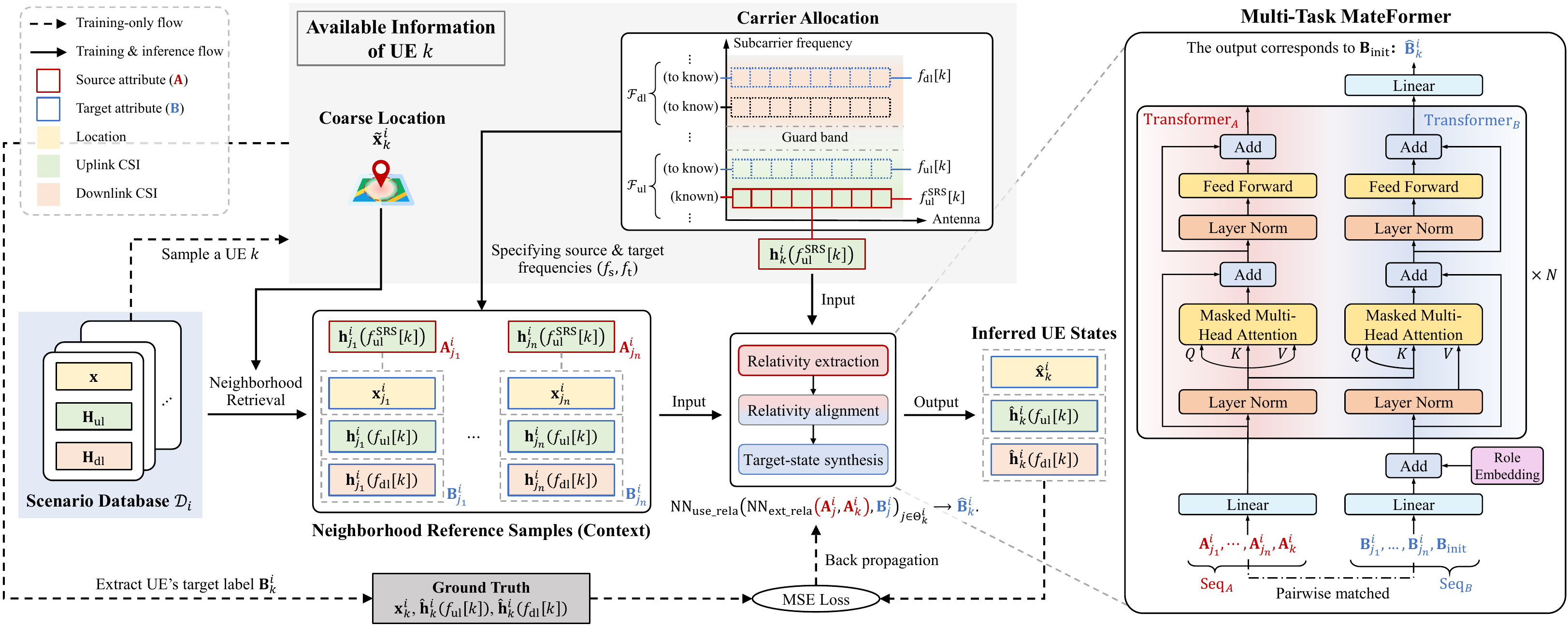}
  \caption{The learning architecture of AWM and the network structure of multi-task MateFormer. AWM achieves performance gains through joint multi-task learning and supports direct deployment in new scenarios after multi-scenario pre-training.}
  \label{fig:awm_architecture}
\vspace{-1em}
\end{figure*}

To learn and use the function in \eqref{eq:relativity_state_transformation}, we design the training and inference architecture of AWM shown in Fig.~\ref{fig:awm_architecture}.
We further propose a multi-task MateFormer as the realization of the WFM, whose network structure is shown on the right side of the figure.

For the model input, we first organize the reference samples in the spatial neighborhood and the source attribute of the current UE into two sequences,
\begingroup
\setlength{\abovedisplayskip}{3pt}
\setlength{\abovedisplayshortskip}{3pt}
\setlength{\belowdisplayskip}{3pt}
\setlength{\belowdisplayshortskip}{3pt}
\begin{align}
\mathrm{Seq}_A
&= [\mathbf A_{j_1}^i,\ldots,\mathbf A_{j_n}^i,\mathbf A_k^i],
\label{eq:sequence_A}\\
\mathrm{Seq}_B
&= [\mathbf B_{j_1}^i,\ldots,\mathbf B_{j_n}^i,\mathbf B_{\mathrm{init}}],
\label{eq:sequence_B}
\end{align}
\endgroup
where $(j_1,\ldots,j_n)$ is an arbitrary ordering of the neighborhood index set $\Theta_k^i$.
The same ordering is used in $\mathrm{Seq}_A$ and $\mathrm{Seq}_B$ to preserve the correspondence between $\mathbf A_j^i$ and $\mathbf B_j^i$ in each reference sample.
The source attribute $\mathbf A_k^i$ of the current UE is appended to $\mathrm{Seq}_A$.
Since its target attribute $\mathbf B_k^i$ is to be inferred, the corresponding position in $\mathrm{Seq}_B$ is filled with a learnable token $\mathbf B_{\mathrm{init}}$.
This token is shared across all channel mapping and wireless localization tasks, and a role embedding distinguishes it from the reference target attributes $\{\mathbf B_j^i\}_{j\in\Theta_k^i}$.
Each CSI is represented as a $2N_{\rm t}$-dimensional real vector by concatenating its real and imaginary parts.
When the target attribute is a location $\mathbf x\in\mathbb R^2$, it is repeated $N_{\rm t}$ times to form the dimension-aligned target-domain input $\mathbf 1_{N_{\rm t}}\otimes\mathbf x\in\mathbb R^{2N_{\rm t}}$.

The multi-task MateFormer consists of $\mathrm{Transformer}_{A}$, an auxiliary Transformer for source-attribute relativity representation, 
and $\mathrm{Transformer}_{B}$, a primary Transformer for relativity-guided target-attribute synthesis.
Each layer employs multi-head attention (MHA) to implement the process formulated in \eqref{eq:relativity_state_transformation}, using attention weights as measures of relativity. 
For layer $\ell$, the core computations are
\begingroup
\setlength{\abovedisplayskip}{3pt}
\setlength{\abovedisplayshortskip}{3pt}
\setlength{\belowdisplayskip}{3pt}
\setlength{\belowdisplayshortskip}{3pt}
\begin{align}
\mathrm{Seq}_A^{(\ell+1)}
&= \mathrm{FFN}_A^{(\ell)}
\left(
\mathrm{MHA}_A^{(\ell)}
(\mathbf Q_A^{(\ell)},\mathbf K_A^{(\ell)},\mathbf V_A^{(\ell)})
\right), \label{eq:mateformer_A}\\
\mathrm{Seq}_B^{(\ell+1)}
&= \mathrm{FFN}_B^{(\ell)}
\left(
\mathrm{MHA}_B^{(\ell)}
(\mathbf Q_A^{(\ell)},\mathbf K_A^{(\ell)},\mathbf V_B^{(\ell)})
\right). \label{eq:mateformer_B}
\end{align}
\endgroup
Specifically, $\mathrm{Transformer}_{B}$ computes attention weights from $\mathbf Q_A^{(\ell)}$ and $\mathbf K_A^{(\ell)}$ projected from $\mathrm{Seq}_A^{(\ell)}$ to evaluate source-domain relativity.
Following the element-wise correspondence between the two sequences, these weights are applied to $\mathbf V_B^{(\ell)}$ projected from $\mathrm{Seq}_B^{(\ell)}$ to guide target-feature synthesis.
Meanwhile, $\mathrm{Transformer}_{A}$ updates the source sequence through self-attention, allowing the next layer to evaluate relativity in a new feature space.
For clarity, the residual connections and layer normalization used in both Transformer branches are omitted from \eqref{eq:mateformer_A} and \eqref{eq:mateformer_B}.
To ensure that relativity evaluation and state synthesis rely strictly on the accurately known reference pairs $\{(\mathbf A_j^i,\mathbf B_j^i)\}_{j\in\Theta_k^i}$, attention masks are applied to $\mathbf A_k^i$ and $\mathbf B_{\mathrm{init}}$ in both sequences. 
By stacking multiple layers of this bipartite structure, MateFormer repeatedly performs relativity extraction, relativity alignment, and target-state synthesis in multiple latent feature spaces.
The output at the position of $\mathbf B_{\mathrm{init}}$ in $\mathrm{Transformer}_{B}$ is mapped to the reconstructed target attribute $\widehat{\mathbf B}_k^i$.
For wireless localization, the output vector is further reshaped into $N_{\rm t}$ two-dimensional vectors and averaged to obtain the location estimate $\widehat{\mathbf x}_k^i$.

Moreover, we define the database of scenario $S_i$ as
\vspace{-0.3em}
\begin{equation}\label{eq:scenario_database}
\vspace{-0.3em}
\mathcal D_i=
\left\{
\left(
\mathbf x_j^i,
\{\mathbf h_j^i(f)\}_{f\in\mathcal F}
\right)
\right\}_{j=1}^{N_i},
\end{equation}
where $\mathcal F=\mathcal F_{\mathrm{ul}}\cup\mathcal F_{\mathrm{dl}}$ is the set of uplink and downlink frequencies in the system.
During training, the system randomly samples an intra-band interpolation, inter-band prediction, or wireless localization task, and selects a data sample from the scenario database $\mathcal D_i$ as the current UE.
Based on the task type, the source attribute $\mathbf A_k^i$ and ground-truth target attribute $\mathbf B_k^i$ are extracted, and neighboring samples are retrieved from $\mathcal D_i$ based on the UE location to construct the reference pairs $\{(\mathbf A_j^i, \mathbf B_j^i)\}_{j\in\Theta_k^i}$.
To reduce numerical scale discrepancies among different attributes, Z-score normalization is applied to locations and CSI at each frequency. 
AWM adopts the mean squared error (MSE) as the unified loss function:
\vspace{-0.5em}
\begin{equation}\label{eq:AWM_loss}
\vspace{-0.5em}
\mathcal L(\boldsymbol\theta)=
\left\|\widehat{\mathbf B}_k^i-\mathbf B_k^i\right\|_2^2,
\end{equation}
where $\mathbf B_k^i$ is the ground-truth location $\mathbf x_k^i$ for wireless localization, and the ground-truth CSI $\mathbf h_k^i(f_{\mathrm{ul}}[k])$ or $\mathbf h_k^i(f_{\mathrm{dl}}[k])$ for channel mapping.
All tasks share the same AWM parameters $\boldsymbol\theta$.
An independent Adam optimizer state is maintained for each task to accommodate different gradient scales and convergence dynamics.
Alternating the above training process among different scenario databases enables multi-scenario joint training.

During deployment, the BS constructs the spatial neighborhood from the coarse UE location $\widetilde{\mathbf{x}}_k^i$ and retrieves the corresponding reference samples from $\mathcal D_i$.
For each channel acquisition or wireless localization task, the source attribute $\mathbf h_k^i(f_{\mathrm{ul}}^{\mathrm{SRS}}[k])$ of the current UE and the corresponding source--target attribute pairs of the reference samples are fed into AWM.
The model then reconstructs either the precise location $\widehat{\mathbf x}_k^i$ or the required target channel, i.e., $\widehat{\mathbf h}_k^i(f_{\mathrm{ul}}[k])$ or $\widehat{\mathbf h}_k^i(f_{\mathrm{dl}}[k])$.
For a continuously moving UE, the initial coarse location is obtained upon initial access via non-neighborhood (random retrieval in \cite{chen2025ALLoc}) AWM localization or via GNSS. The former requires no prior location information.
Thereafter, the current location estimate serves as the coarse location for the next time slot.
This establishes a self-sustaining closed-loop workflow:
coarse localization $\to$ spatial-neighborhood retrieval $\to$ reference-pair construction $\to$ localization and channel mapping $\to$ location reuse as the next-slot coarse location.

\section{Experiments and Results} \label{sec:experiments}

\begin{table*}[!t]\scriptsize
\renewcommand{\arraystretch}{1.5}
\caption{\small Performance comparison of channel mapping and localization under different training schemes in the single scenario.}
\vspace{-0.5em}
\setlength{\tabcolsep}{3.5pt}
\vspace{-0.8em}
\begin{center}
\begin{tabular}{C{2.55cm}|C{1.05cm}|C{1.05cm}|C{1.88cm}C{1.88cm}C{1.72cm}C{1.88cm}C{1.88cm}C{1.72cm}}
\hline
\multicolumn{1}{@{}c@{}|}{\multirow{2}{*}{\diagbox[width=\dimexpr 2.55cm+2\tabcolsep\relax, height=1.82\baselineskip]{Tasks}{Schemes}}} & \multirow[c]{2}{1.05cm}{\centering AWM} & \multirow[c]{2}{1.05cm}{\centering\makecell{AWM\\(single \\optim)}} & \multicolumn{3}{c|}{AWM trained on specific task and frequency} & \multicolumn{3}{c}{ResMLP (only support specific task and frequency)} \\ \cline{4-9} 
& & & \multicolumn{1}{c|}{\scriptsize $5^\text{th}$-up to $24^\text{th}$-up} & \multicolumn{1}{c|}{\scriptsize $5^\text{th}$-up to $24^\text{th}$-down} & \multicolumn{1}{c|}{\scriptsize $5^\text{th}$-up to loc} & \multicolumn{1}{c|}{\scriptsize $5^\text{th}$-up to $24^\text{th}$-up} & \multicolumn{1}{c|}{\scriptsize $5^\text{th}$-up to $24^\text{th}$-down}& \multicolumn{1}{c}{\scriptsize $5^\text{th}$-up to loc} \\ \hline
Intra-band interpolation & -18.57\,\text{dB} & -17.26\,\text{dB} & \multicolumn{1}{c|}{-9.77\,\text{dB}} & \multicolumn{1}{c|}{---} & \multicolumn{1}{c|}{---} & \multicolumn{1}{c|}{-1.01\,\text{dB}} & \multicolumn{1}{c|}{---} & \multicolumn{1}{c}{---} \\ \hline
Inter-band prediction & -10.43\,\text{dB} & -10.76\,\text{dB} & \multicolumn{1}{c|}{---} & \multicolumn{1}{c|}{1.28\,\text{dB}} & \multicolumn{1}{c|}{---} & \multicolumn{1}{c|}{---} & \multicolumn{1}{c|}{1.59\,\text{dB}} & \multicolumn{1}{c}{---} \\ \hline
Uplink localization & 0.0949\,\text{m} & 0.2153\,\text{m} & \multicolumn{1}{c|}{---} & \multicolumn{1}{c|}{---} & \multicolumn{1}{c|}{0.2991\,\text{m}} & \multicolumn{1}{c|}{---} & \multicolumn{1}{c|}{---} & \multicolumn{1}{c}{1.9569\,\text{m}} \\ \hline
\end{tabular}
\end{center}
\vspace{-1em}
\label{tab:single_scenario_performance}
\end{table*}

\begin{table*}[!t]\scriptsize
\renewcommand{\arraystretch}{1.3}
\renewcommand{\cellset}{\renewcommand{\arraystretch}{1.0}}
\caption{\small Cross-scenario generalization performance of AWM with multi-scenario pre-training.}
\vspace{-0.5em}
\setlength{\tabcolsep}{2.5pt}
\vspace{-0.8em}
\begin{center}
\begin{tabular}{C{2.55cm}|C{2.1cm}|C{2.1cm}|C{2.1cm}|C{1.95cm}|C{1.85cm}C{1.95cm}C{1.5cm}}
\hline
\multicolumn{1}{@{}c@{}|}{\multirow{2}{*}{\diagbox[width=\dimexpr 2.55cm+2\tabcolsep\relax, height=1.98\baselineskip]{Tasks}{Schemes}}} & \multirow{2.4}{*}{\makecell{Pre-trained AWM\\(6.38M, testing on\\a learned scenario)}} & \multirow{2.4}{*}{\makecell{Pre-trained AWM\\(6.38M, testing\\on new scenario)}} & \multirow{2.4}{*}{\makecell{Pre-trained AWM\\(25.4M, testing\\on new scenario)}} & \multirow{2.4}{*}{\makecell{Scenario-specific\\AWM\\(6.38M)}} & \multicolumn{3}{c}{\makecell{Scenario-specific ResMLP\\(only support specific task and frequency)}} \\ \cline{6-8}
& & & & & \multicolumn{1}{c|}{\scriptsize $5^\text{th}$-up to $24^\text{th}$-up} & \multicolumn{1}{c|}{\scriptsize $5^\text{th}$-up to $24^\text{th}$-down} & \multicolumn{1}{c}{\scriptsize $5^\text{th}$-up to loc} \\ \hline
Intra-band interpolation & -21.80\,\text{dB} & -20.51\,\text{dB} &  -24.44\,\text{dB} &  -22.84\,\text{dB} & \multicolumn{1}{c|}{-0.76\,\text{dB}} & \multicolumn{1}{c|}{---} & \multicolumn{1}{c}{---} \\ \hline
Inter-band prediction & -12.83\,\text{dB} & -12.35\,\text{dB} &  -16.14\,\text{dB} &  -14.01\,\text{dB} & \multicolumn{1}{c|}{---} & \multicolumn{1}{c|}{1.26\,\text{dB}} & \multicolumn{1}{c}{---} \\ \hline
Uplink localization & 0.0651\,\text{m} & 0.0758\,\text{m} &  0.0470\,\text{m} &  0.0552\,\text{m} & \multicolumn{1}{c|}{---} & \multicolumn{1}{c|}{---} & \multicolumn{1}{c}{1.9543\,\text{m}} \\ \hline
\end{tabular}
\end{center}
\label{tab:multi_scenario_performance}
\vspace{-2em}
\end{table*}

\subsection{Experiment Settings} \label{subsec:exp_settings}

We evaluate the performance on the open-source ray-tracing simulation dataset DeepMIMO \cite{Alkhateeb2019DeepMIMO}, including the single scenario `O1' and the multi-scenario `MO1' consisting of five adjacent cells \cite{chen2025ALLoc}.
Each BS is equipped with a ULA of $N_{\rm t} = 32$ antennas.
The uplink and downlink center frequencies are set to $3.5\text{ GHz}$ and $3.6\text{ GHz}$ with a bandwidth of $40\text{ MHz}$, respectively, each divided into $N_{\rm c} = 32$ subcarriers.
In each scenario, $20{,}000$ training samples and $20{,}000$ test samples are generated.
For the neighborhood construction in AWM, $16$ to $64$ samples are randomly sampled from the UE's spatial neighborhood as reference samples during training, while $64$ nearest reference samples are retrieved based on the coarse UE location during testing.
During testing, without loss of generality, 
the 5th subcarrier of the uplink band is designated as the default SRS sounding carrier, and the 24th subcarriers of the uplink and downlink bands serve as the target carriers for intra-band channel interpolation and inter-band channel prediction, respectively.
To fully reflect the generalization of AWM in unseen frequencies, the default training of AWM excludes these three subcarriers.

For the MateFormer network in AWM, the model depth is set to $8$, the number of attention heads is set to $8$, and the default hidden dimension size is $256$.
Different parameter scales of AWM variants can be constructed by adjusting this feature dimension.
Existing wireless (foundation) models typically require different weights for different tasks, frequencies, and scenarios, making direct comparisons with AWM difficult. To establish a meaningful baseline, we employ a dedicated model for each task and scenario and compare their collective performance with AWM.
Specifically, we use the residual multilayer perceptron (ResMLP), widely used in channel mapping and wireless localization \cite{yang2019MLPFDD, Artan2021DNNLoc}, to construct 15 task- and scenario-specific baselines. Under the default configurations, AWM contains $6.38\text{ M}$ parameters and requires training only once, whereas each ResMLP contains $2.14\text{ M}$ parameters, totaling $32.1\text{ M}$ across 15 separately trained models. Also, since ResMLP lacks cross-frequency generalization capability, we train it on the specific subcarriers used in the testing tasks.
We adopt the normalized mean squared error (NMSE) as the evaluation metric for channel acquisition, and the mean absolute error (MAE) for wireless localization, which are defined as
\vspace{-0.3em}
\begin{equation}\label{eq:NMSE}
\vspace{-0.5em}
    \mathrm{NMSE} = \mathbb{E} \left\{ \|\widehat{\mathbf{h}} - \mathbf{h}\|_2^2 \big/ \|\mathbf{h}\|_2^2 \right\}
\end{equation}
\vspace{-0.8em}
\begin{equation} \label{eq:MAE}
\vspace{-0.5em}
    \mathrm{MAE} = \mathbb{E} \left\{ \left\| \widehat{\mathbf{x}} - \mathbf{x} \right\|_2 \right\},
\end{equation}
where $\mathbf{h}$ and $\widehat{\mathbf{h}}$ denote the ground-truth and estimated CSI, respectively, and $\mathbf{x}$ and $\widehat{\mathbf{x}}$ denote the ground-truth and estimated UE locations, respectively.

\subsection{Performance Evaluation} \label{subsec:results}

We first evaluate AWM on the three wireless tasks in a specific scenario `O1'.
As shown in Table~\ref{tab:single_scenario_performance}, when trained on specific tasks and frequencies, AWM consistently outperforms the ResMLP baseline, demonstrating the benefits of advanced model structure and the reference information provided by neighboring samples.
Under the proposed multi-task learning framework, AWM operates effectively on unseen testing subcarriers, significantly improving individual task performance.
In particular, multi-task training exhibits \textit{emergent abilities} in inter-band channel prediction \cite{wei2022emergent}, reducing the NMSE from an unusable regime of $1.28\text{ dB}$ to a practical level of $-10.76\text{ dB}$.
Furthermore, the multi-optimizer scheme adaptively balances the optimization dynamics among distinct tasks, reducing the localization MAE from $0.2153\text{ m}$ to $0.0949\text{ m}$ and further improving the overall performance of AWM.

\begin{figure}[!t]
\centering
\includegraphics[width=\linewidth]{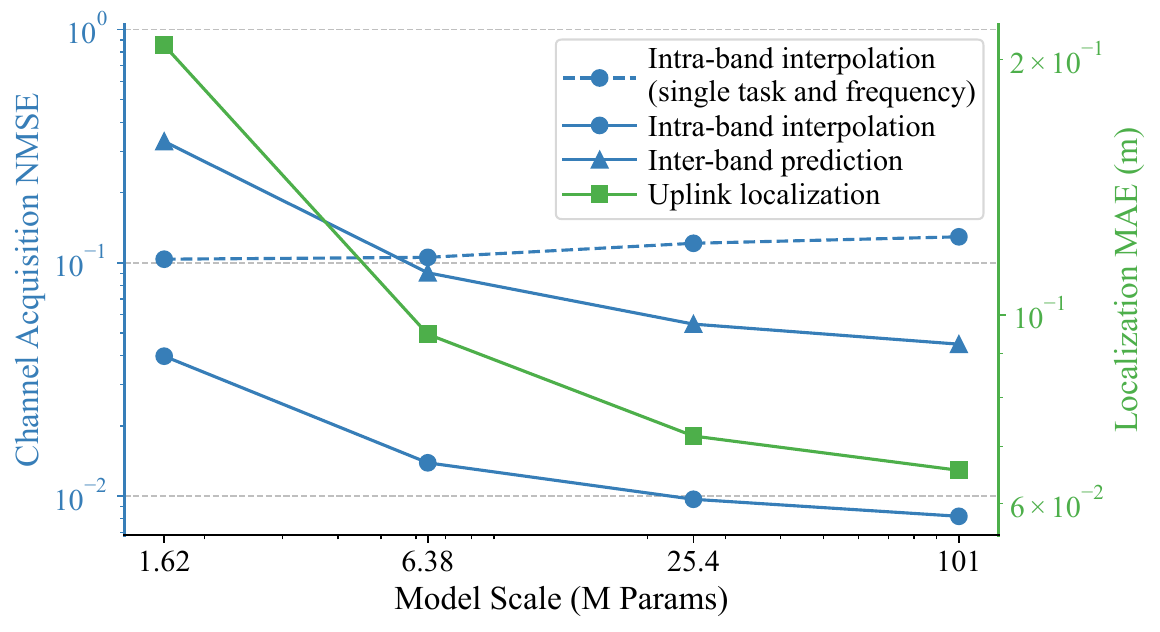}
\vspace{-1.5em}
\caption{\small Performance of AWM under different model parameter scales for channel mapping and localization tasks.}
\label{fig:scaling}
\vspace{-1em}
\end{figure}

Fig.~\ref{fig:scaling} illustrates the impact of model parameter scales on the performance of AWM in scenario `O1'.
Under the multi-task learning mode, as the model scale increases from $1.62\text{ M}$ to $101\text{ M}$, the reconstruction errors of AWM in intra-band interpolation, inter-band prediction, and wireless localization all exhibit a steady downward trend.
In contrast, scaling AWM trained solely on intra-band channel interpolation does not yield a consistent performance improvement.
This further indicates that unified multi-task learning increases the information density of the training data and thus lays the foundation for wireless model's scaling behaviors in parameter amount dimension.

\begin{figure}[!t]
  \centering
  \captionsetup{skip=2pt}
  \begin{minipage}{0.90\linewidth}
    \centering
    \includegraphics[width=\linewidth]{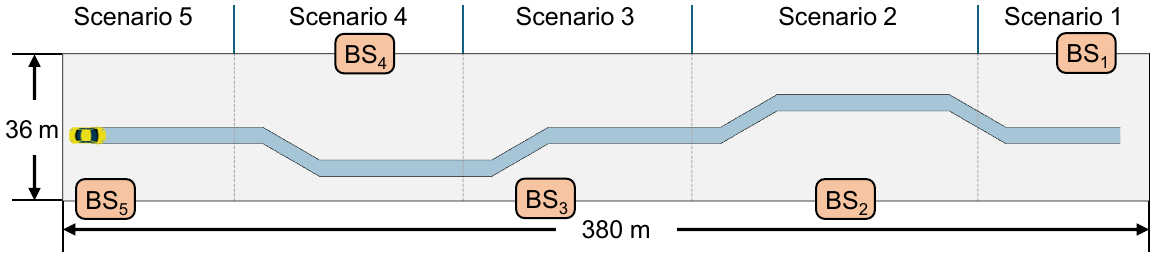}
    \caption*{\footnotesize (a) The trajectory of a user's movement in `MO1'. During this movement (1800 time slots), the user moves through five scenarios and occupies different subcarriers for uplink and downlink access in each scenario.}
  \end{minipage}
  \par\smallskip
  \begin{minipage}{0.90\linewidth}
    \centering
    \includegraphics[width=\linewidth]{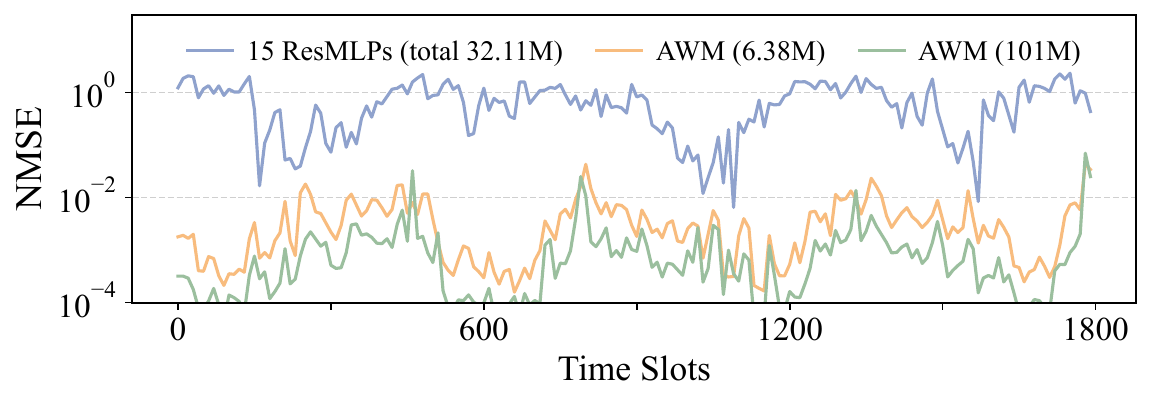}
    \caption*{\footnotesize (b) Intra-band channel interpolation performance during the movement.}
  \end{minipage}
  \par\smallskip
  \begin{minipage}{0.90\linewidth}
    \centering
    \includegraphics[width=\linewidth]{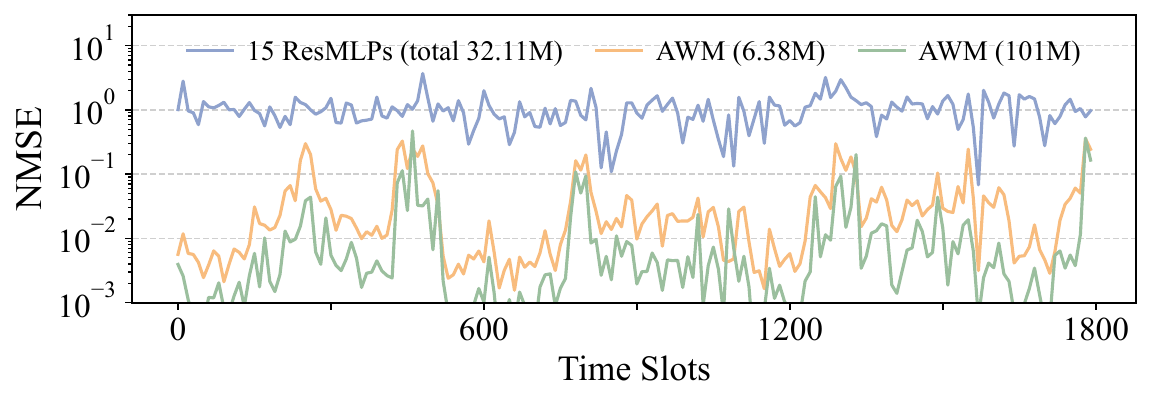}
    \caption*{\footnotesize (c) Inter-band channel prediction performance during the movement.}
  \end{minipage}
  \par\smallskip
  \begin{minipage}{0.90\linewidth}
    \centering
    \includegraphics[width=\linewidth]{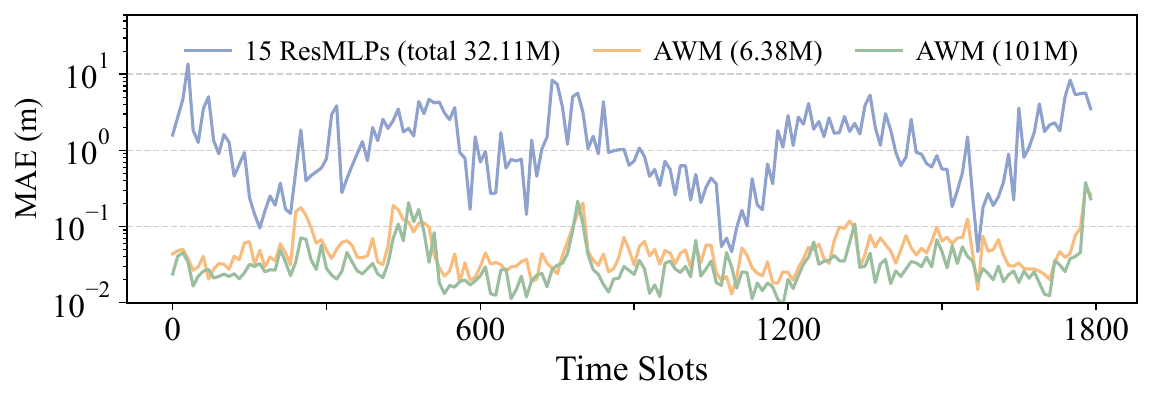}
    \caption*{\footnotesize (d) User localization performance during the movement.}
  \end{minipage}
  \par
  \vspace{0.5em}
  \caption{\small Continuously serve a mobile user through AWM or $15$ task- and scenario-specific ResMLPs.
  A moving average filter with a window size of 11 is applied for smoothing.}
\vspace{-1.2em}
\label{fig:trace_application}
\end{figure}

To further evaluate cross-scenario generalization, we pre-train AWM on four cells in `MO1' and directly test it on a held-out cell, as summarized in Table~\ref{tab:multi_scenario_performance}.
The results show that the pre-trained AWM can be directly deployed in an unseen scenario, with performance close to that on the learned scenarios across all tasks, demonstrating the cross-scenario generalization inherited from AL \cite{chen2025ALLoc}.
Consistent with the scaling trend in Fig.~\ref{fig:scaling}, multi-task learning increases the information density of the training data, leaving the models underfitted.
As more scenarios are jointly learned, the model capacity is preferably scaled proportionally with the training data; after scaling, AWM outperforms the model trained specifically for the target scenario in the unseen scenario, thereby fully realizing the benefit of pre-training.
Finally, we evaluate continuous service for a moving UE to emulate the deployment of AWM in a practical system.
As shown in Fig.~\ref{fig:trace_application}(a), the UE travels through five cells in `MO1' over $1800$ time slots, with dynamically varying uplink and downlink subcarrier allocation.
At each time slot, AWM uses the previous localization result as the coarse location to retrieve reference samples and then performs channel mapping and localization, as described in Section~\ref{sec:AWM}.
Figs.~\ref{fig:trace_application}(b)--(d) show that AWM reduces average NMSE by approximately two orders of magnitude for intra-band interpolation and more than one order for inter-band prediction compared with the ResMLPs.
The mean localization error decreases from the meter level to the centimeter level.
All three tasks operate throughout the trajectory without error-accumulation-induced divergence.
Compared to conventional schemes that require training and deploying $15$ scenario- and task-specific ResMLP models, AWM requires only one model while providing superior wireless services, enabling a highly simplified intelligent system.


\section{Conclusion} \label{conclusion}
In this paper, we presented AWM as a step toward unified wireless intelligence. A single model with a single set of weights supports diverse communication and sensing tasks across scenarios and frequency bands, while cross-task and cross-frequency learning yields emergent gains and enables continuous closed-loop operation. These advances are driven by a unified relativity representation paradigm, analogous to next-token prediction in language models, which casts heterogeneous wireless functions into a foundation framework and significantly reuses the training data, model capacity, and learning cost. Implemented through AL, this paradigm also exhibits clear scaling behaviors.
We hope this work can help pave the way toward a ``ChatGPT moment'' for wireless AI.


\section*{Acknowledgment}
This work was supported in part by National Natural Science Foundation of China under Grant 62394292 and 62631025, 
and in part by the Fundamental Research Funds for the Central Universities under Grant 226-2024-00069. 
The authors Zirui Chen and Ziqing Xing have equal contribution to this work. 
(\textit{Corresponding Author: Zhaoyang~Zhang})


\small

\bibliographystyle{IEEEtran}
\bibliography{IEEEabrv.bib, myabrv.bib, ref.bib}

\end{document}